\documentclass[aps,prl,superscriptaddress,showpacs,floatfix,twocolumn]{revtex4}
\usepackage{amsmath}
\ifx\pdftexversion\undefined
  \usepackage[dvips]{graphicx}
\else
  \usepackage[pdftex]{graphicx}
\fi

\newcommand{\pp} {\mbox{$p$$+$$p$}}
\newcommand{\pt} {\mbox{$p_\perp$}}
\newcommand{\piz} {\mbox{$\pi^0$}}
\newcommand{\pbar} {\mbox{$\overline{p}$}}
\newcommand{\snn} {\mbox{$\sqrt{s_{NN}}$}}
\newcommand{\npart} {\mbox{$N_{\it part}$}}
\newcommand{\ncoll} {\mbox{$N_{\it coll}$}}

\def\Journal#1#2#3#4{{#1}{\bf #2}, #3 (#4)}

\def\NIMA{{Nucl. Instrum. Methods}~{\bf A}}
\def\NPA{{Nucl. Phys.}~{\bf A}}
\def\NPB{{Nucl. Phys.}~{\bf B}}
\def\PLB{{Phys. Lett.}~{\bf B}}

\def\PRL{Phys. Rev. Lett.\ }
\def\PRD{{Phys. Rev.}~{\bf D}}
\def\PRC{{Phys. Rev.}~{\bf C}}

\def\ZPC{{Z. Phys.}~{\bf C}}

\begin{document}

\title{Scaling properties of proton and anti-proton production \\
in $\snn=200$\,GeV Au$+$Au collisions}

\newcommand{\abilene}{Abilene Christian University, Abilene, TX 79699, USA}
\newcommand{\acadsin}{Institute of Physics, Academia Sinica, Taipei 11529, Taiwan}
\newcommand{\banaras}{Department of Physics, Banaras Hindu University, Varanasi 221005, India}
\newcommand{\barc}{Bhabha Atomic Research Centre, Bombay 400 085, India}
\newcommand{\bnl}{Brookhaven National Laboratory, Upton, NY 11973-5000, USA}
\newcommand{\caucr}{University of California - Riverside, Riverside, CA 92521, USA}
\newcommand{\ciae}{China Institute of Atomic Energy (CIAE), Beijing, People's Republic of China}
\newcommand{\cns}{Center for Nuclear Study, Graduate School of Science, University of Tokyo, 7-3-1 Hongo, Bunkyo, Tokyo 113-0033, Japan}
\newcommand{\columbia}{Columbia University, New York, NY 10027 and Nevis Laboratories, Irvington, NY 10533, USA}
\newcommand{\dapnia}{Dapnia, CEA Saclay, F-91191, Gif-sur-Yvette, France}
\newcommand{\debrecen}{Debrecen University, H-4010 Debrecen, Egyetem t{\'e}r 1, Hungary}
\newcommand{\fsu}{Florida State University, Tallahassee, FL 32306, USA}
\newcommand{\gsu}{Georgia State University, Atlanta, GA 30303, USA}
\newcommand{\hiroshima}{Hiroshima University, Kagamiyama, Higashi-Hiroshima 739-8526, Japan}
\newcommand{\ihepprot}{Institute for High Energy Physics (IHEP), Protvino, Russia}
\newcommand{\isu}{Iowa State University, Ames, IA 50011, USA}
\newcommand{\jinrdubna}{Joint Institute for Nuclear Research, 141980 Dubna, Moscow Region, Russia}
\newcommand{\kaeri}{KAERI, Cyclotron Application Laboratory, Seoul, South Korea}
\newcommand{\kangnung}{Kangnung National University, Kangnung 210-702, South Korea}
\newcommand{\kek}{KEK, High Energy Accelerator Research Organization, Tsukuba-shi, Ibaraki-ken 305-0801, Japan}
\newcommand{\kfki}{KFKI Research Institute for Particle and Nuclear Physics (RMKI), H-1525 Budapest 114, POBox 49, Hungary}
\newcommand{\korea}{Korea University, Seoul, 136-701, Korea}
\newcommand{\kurchatov}{Russian Research Center ``Kurchatov Institute", Moscow, Russia}
\newcommand{\kyoto}{Kyoto University, Kyoto 606, Japan}
\newcommand{\labllr}{Laboratoire Leprince-Ringuet, Ecole Polytechnique, CNRS-IN2P3, Route de Saclay, F-91128, Palaiseau, France}
\newcommand{\lawllnl}{Lawrence Livermore National Laboratory, Livermore, CA 94550, USA}
\newcommand{\losalamos}{Los Alamos National Laboratory, Los Alamos, NM 87545, USA}
\newcommand{\lpc}{LPC, Universit{\'e} Blaise Pascal, CNRS-IN2P3, Clermont-Fd, 63177 Aubiere Cedex, France}
\newcommand{\lund}{Department of Physics, Lund University, Box 118, SE-221 00 Lund, Sweden}
\newcommand{\muenster}{Institut fuer Kernphysik, University of Muenster, D-48149 Muenster, Germany}
\newcommand{\myongji}{Myongji University, Yongin, Kyonggido 449-728, Korea}
\newcommand{\nagasaki}{Nagasaki Institute of Applied Science, Nagasaki-shi, Nagasaki 851-0193, Japan}
\newcommand{\newmex}{University of New Mexico, Albuquerque, NM, USA}
\newcommand{\nmsu}{New Mexico State University, Las Cruces, NM 88003, USA}
\newcommand{\ornl}{Oak Ridge National Laboratory, Oak Ridge, TN 37831, USA}
\newcommand{\orsay}{IPN-Orsay, Universite Paris Sud, CNRS-IN2P3, BP1, F-91406, Orsay, France}
\newcommand{\pnpi}{PNPI, Petersburg Nuclear Physics Institute, Gatchina, Russia}
\newcommand{\riken}{RIKEN (The Institute of Physical and Chemical Research), Wako, Saitama 351-0198, JAPAN}
\newcommand{\rkrbrc}{RIKEN BNL Research Center, Brookhaven National Laboratory, Upton, NY 11973-5000, USA}
\newcommand{\saispbstu}{St. Petersburg State Technical University, St. Petersburg, Russia}
\newcommand{\saopaulo}{Universidade de S{\~a}o Paulo, Instituto de F\'{\i}sica, Caixa Postal 66318, S{\~a}o Paulo CEP05315-970, Brazil}
\newcommand{\seoulnat}{System Electronics Laboratory, Seoul National University, Seoul, South Korea}
\newcommand{\stonybrkc}{Chemistry Department, Stony Brook University, SUNY, Stony Brook, NY 11794-3400, USA}
\newcommand{\stonycrkp}{Department of Physics and Astronomy, Stony Brook University, SUNY, Stony Brook, NY 11794, USA}
\newcommand{\subatech}{SUBATECH (Ecole des Mines de Nantes, CNRS-IN2P3, Universit{\'e} de Nantes) BP 20722 - 44307, Nantes, France}
\newcommand{\tenn}{University of Tennessee, Knoxville, TN 37996, USA}
\newcommand{\titech}{Department of Physics, Tokyo Institute of Technology, Tokyo, 152-8551, Japan}
\newcommand{\tsukuba}{Institute of Physics, University of Tsukuba, Tsukuba, Ibaraki 305, Japan}
\newcommand{\vandy}{Vanderbilt University, Nashville, TN 37235, USA}
\newcommand{\waseda}{Waseda University, Advanced Research Institute for Science and Engineering, 17 Kikui-cho, Shinjuku-ku, Tokyo 162-0044, Japan}
\newcommand{\weizmann}{Weizmann Institute, Rehovot 76100, Israel}
\newcommand{\yonsei}{Yonsei University, IPAP, Seoul 120-749, Korea}
\affiliation{\abilene}
\affiliation{\acadsin}
\affiliation{\banaras}
\affiliation{\barc}
\affiliation{\bnl}
\affiliation{\caucr}
\affiliation{\ciae}
\affiliation{\cns}
\affiliation{\columbia}
\affiliation{\dapnia}
\affiliation{\debrecen}
\affiliation{\fsu}
\affiliation{\gsu}
\affiliation{\hiroshima}
\affiliation{\ihepprot}
\affiliation{\isu}
\affiliation{\jinrdubna}
\affiliation{\kaeri}
\affiliation{\kangnung}
\affiliation{\kek}
\affiliation{\kfki}
\affiliation{\korea}
\affiliation{\kurchatov}
\affiliation{\kyoto}
\affiliation{\labllr}
\affiliation{\lawllnl}
\affiliation{\losalamos}
\affiliation{\lpc}
\affiliation{\lund}
\affiliation{\muenster}
\affiliation{\myongji}
\affiliation{\nagasaki}
\affiliation{\newmex}
\affiliation{\nmsu}
\affiliation{\ornl}
\affiliation{\orsay}
\affiliation{\pnpi}
\affiliation{\riken}
\affiliation{\rkrbrc}
\affiliation{\saispbstu}
\affiliation{\saopaulo}
\affiliation{\seoulnat}
\affiliation{\stonybrkc}
\affiliation{\stonycrkp}
\affiliation{\subatech}
\affiliation{\tenn}
\affiliation{\titech}
\affiliation{\tsukuba}
\affiliation{\vandy}
\affiliation{\waseda}
\affiliation{\weizmann}
\affiliation{\yonsei}
\author{S.S.~Adler}	\affiliation{\bnl}
\author{S.~Afanasiev}	\affiliation{\jinrdubna}
\author{C.~Aidala}	\affiliation{\bnl}
\author{N.N.~Ajitanand}	\affiliation{\stonybrkc}
\author{Y.~Akiba}	\affiliation{\kek} \affiliation{\riken}
\author{J.~Alexander}	\affiliation{\stonybrkc}
\author{R.~Amirikas}	\affiliation{\fsu}
\author{L.~Aphecetche}	\affiliation{\subatech}
\author{S.H.~Aronson}	\affiliation{\bnl}
\author{R.~Averbeck}	\affiliation{\stonycrkp}
% ...
\collaboration{PHENIX Collaboration} \noaffiliation

\date{\today}        
\begin{abstract}
We report on the yield of protons and anti-protons, as a function of
centrality and transverse momentum, in Au$+$Au collisions
at $\snn$ =  200 GeV measured at mid-rapidity by the PHENIX experiment 
at RHIC.  In central collisions at intermediate transverse momenta 
($1.5 < \pt < 4.5$\,GeV/$c$) a significant fraction of all produced 
particles are protons and anti-protons. They show a centrality-scaling 
behavior different from that of pions.  The $\pbar/\pi$ and $p/\pi$ ratios are 
enhanced compared to peripheral Au+Au, p+p and $e^{+}e^{-}$ collisions. 
This enhancement is limited to $\pt < 5$\,GeV/$c$ as deduced from the 
ratio of charged hadrons to $\piz$ measured in the range 
$1.5 < \pt < 9$\,GeV/$c$.
\end{abstract}

\pacs{25.75.Dw}
\maketitle

Heavy-ion collisions at RHIC energies permit the study of 
nuclear matter at extreme energy densities.  Hadrons originating from
fragmentation of partons that have undergone 
large momentum transfer (hard) scatterings are sensitive 
probes of the hottest and densest 
stage of the collision.  Data collected during the  
first RHIC run at$\snn$ = 130 GeV led to the discovery of suppression of 
high transverse momentum ($\pt \ge 2$\,GeV/$c$) hadron production in central 
Au$+$Au collisions~\cite{ppg003,ppg013,star_suppr} when compared to 
expectations from nucleon-nucleon collisions. This effect, quantified in 
terms of a nuclear modification factor 
$R_{\rm AA}= \left(yield_{\rm AA} / N_{\it coll}\right)/ yield_{pp}$,
where $N_{\it coll}$ is the average number of binary nucleon-nucleon collisions,
had been discussed as a possible consequence of the energy loss
suffered by partons moving in a dense medium~\cite{quench_effect,quenching_theory}. 
Unexpectedly, it was found that $R_{\rm AA}$ is more strongly suppressed 
for $\pi^0$ than for charged hadrons~\cite{ppg003}, and that 
the yields of $p$ and $\pbar$ near 2\,GeV/$c$ in central 
events~\cite{ppg006} are comparable to the yield of pions ($p/\pi\sim 1$). 
This is in contrast to the $p/\pi$ 
ratios of $\sim$ 0.1 - 0.3, measured in $\pp$~\cite{ISR} and 
$e^{+}e^{-}$~\cite{DELPHI} collisions, and to perturbative 
quantum chromodynamics phenomenology~\cite{ptopi_bjunctions}.
These results suggest that an investigation of particle composition 
is important for understanding the medium effect on high-$\pt$
phenomena at RHIC. During the 2001 Au$+$Au run at $\snn= 200$\,GeV the PHENIX 
experiment collected data to study  
the scaling properties of $p$ and $\pbar$ production as well as 
the $p/\pi$, $\pbar/\pi$ and charged hadron to pion ($h/\pi$) ratios as 
a function of centrality.

The PHENIX detector~\cite{NIM} combines high momentum resolution 
with diverse particle identification (PID), resulting in 
hadron identification over a broad momentum range.
The present results combine the measurements of
$\pi^{\pm}$, $p$ and $\pbar$ with
those of neutral pions~\cite{ppg014} and inclusive 
charged hadrons~\cite{JiaQM02}.  A ``minimum bias'' trigger based on
signals from the Beam-Beam Counters (BBC) and Zero-Degree Calorimeters
(ZDC) sampled $92.2^{+2.5}_{-3}\%$ of the inelastic Au$+$Au 
cross-section of $\sigma_{inel}^{AuAu}$~=~$6.9$\,b~\cite{ppg014}. 
The collision vertex is restricted to
$\pm$ 30\,cm of the nominal origin.  Approximately $2\times 10^7$
($3\times 10^7$) minimum bias events are used in the charged
(neutral) particle analysis. These samples are subdivided into 7
centrality classes based on cuts in the combined ZDC and BBC response:
0-10\%, 10-20\%, 20-30\%, 30-40\%, 40-50\%, 50-60\%, 60-92\% 
of $\sigma_{inel}^{AuAu}$.
The average number of participants ($\npart$) and collisions $\ncoll$
for each centrality class are derived from a Glauber model calculation~\cite{ppg014}.

Identified charged particles are measured over a subset of the PHENIX 
East-arm spectrometer  covering pseudo-rapidity $\lvert\eta\rvert< 0.35$ and
$\Delta\phi=\pi/8$ in azimuthal angle.
PID is based on particle mass calculated from the measured 
momentum and velocity.  The momentum resolution 
is $\delta p/p \simeq 0.7\% \oplus 1\%\times p$\,(GeV/$c$)
and is provided by a
multi-layer drift chamber (DC) followed by a multi-wire proportional 
chamber with pad readout (PC1).  The velocity is obtained by measuring
the time-of-flight (TOF) and the path length along the trajectory.  The
timing system uses the BBC to provide a global start signal; hits on
the TOF scintillator wall, located at a radial
distance of 5.06\,m, provide individual stop signals.  The 
resolution is $\sigma\simeq115$\,ps, which allows a 4$\sigma$ $\pi/K$ and 
$K/p$ separation up to $\pt \simeq 2$\,GeV/$c$ and $\pt \simeq 4$\,GeV/$c$, 
respectively. A $2\sigma$ momentum dependent cut in mass squared is used up 
to  $\pt = 2$\,GeV/$c$ and $\pt =4$\,GeV/$c$ to select $\pi$ and $(\pbar)p$. 
Asymmetric cuts are applied at higher momenta to extend the 
$\pi/K$ and $K/p$ separation up to $\pt$ of 3 and 4.5\,GeV/$c$.
The spectra are corrected for geometrical acceptance, decay-in-flight, and
reconstruction efficiency using a GEANT-based Monte Carlo
(MC) simulation and embedding simulated tracks into real events with
different particle multiplicities.

The $p$ and $\pbar$ yields are corrected for feed-down from weak decays 
using a MC simulation and the measured
$\Lambda/p$ and $\overline{\Lambda}/\pbar$ ratios at $\snn = 130$
GeV~\cite{lambda} which include contributions from $\Xi$ and $\Sigma^{0}$.   
Corrections for feed-down from $\Sigma^{\pm}$ are not applied, 
but estimates based on HIJING MC give less than $\sim 5$\% 
contribution. At $\pt = 0.65$\,GeV/$c$, about 40\% of the inclusive 
$(\pbar)p$ come from weak decays. This fraction reduces to $\approx 25$\% 
at 4\,GeV/$c$. The systematic uncertainty of this correction is
estimated at 6\% by varying the $\Lambda/p$  ($\overline{\Lambda}/\pbar$) 
ratios within the $\pm 24\%$ errors of the  $\snn = 130$ measurement
and assuming $m_{T}$-scaling at high-$\pt$. The above 
uncertainty  could be larger if  the $\Lambda/p $ ($\overline{\Lambda}/\pbar$) 
ratios change significantly with $\pt$ and beam energy.
The additional systematic error on the overall 
normalization is 8\% for $\pt <3 $\,GeV/$c$ and 12\% above 3\,GeV/$c$. 
Added in quadrature, the total systematic errors are 11\% and 14\%; 
the larger value is for $\pt > 3$\,GeV/$c$. 

Inclusive charged hadrons are measured in the West-arm spectrometer
 covering $\lvert\eta\rvert< 0.35$ and $\Delta\phi=\pi/2$. 
Two pad chambers (PC2, PC3) located at 4.2\,m and 5\,m,
respectively and a Ring Imaging $\check{\rm C}$erenkov
Counter~\cite{JiaQM02} are used to reject and
subtract high-$\pt$ background.
The systematic error on the yields range 
from 11\%
for $\pt<5$\,GeV/$c$ to 45\% at 9 GeV/$c$.

Neutral pions are reconstructed via the decay $\piz \rightarrow
\gamma\gamma$ through an invariant mass analysis of $\gamma$ pairs
detected in the electromagnetic calorimeter (EMCal), which covers
$\Delta\eta = 0.7$ and $\Delta \phi = \pi$.  The absolute energy scale
is known to $\leq$~1.5\%.  The systematic errors on the $\piz$ spectra
range from 10\% to 17\%, from low to high $\pt$~\cite{ppg014}.

\begin{figure}[tb]
\center
\includegraphics[width=1.0\linewidth]{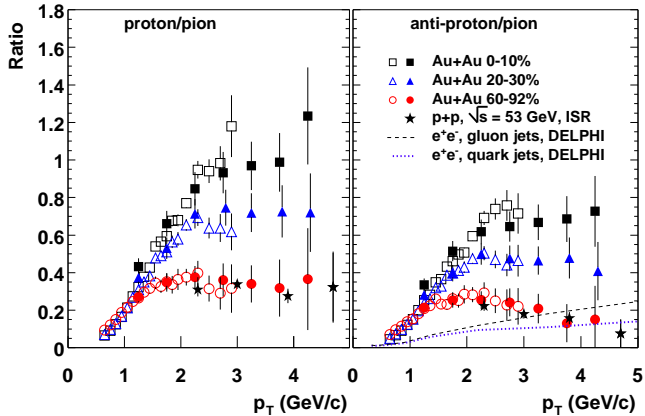}
\caption[]{$p/\pi$ (left) and $\pbar/\pi$ ratios for central(0-10\%), 
mid-central(20-30\%) and peripheral (60-92\%) Au+Au collisions at 
$\snn = 200$ GeV. Open (filled) points are for 
$\pi^{\pm}$ ($\pi^0$), respectively. Data from 
$\sqrt{s} = 53 $ GeV $\pp$ collisions~\cite{ISR} are shown with stars. 
The dashed and dotted lines are $(\pbar+p)/(\pi^{+}+\pi^{-})$ ratio in 
gluon and in quark jets~\cite{DELPHI}.}
\label{fig:1}
\end{figure}

Figure~\ref{fig:1} shows the $p/\pi$ and $\pbar/\pi$ ratios as a
function of $\pt$ measured at mid-rapidity in central (0--10\%),
mid-central (20--30\%), and peripheral (60--92\%) Au$+$Au collisions.
The open symbols represent the $p/\pi^{+}$ and $\pbar/\pi^{-}$
measurements, while the closed symbols represent the corresponding
$p/\piz$ and $\pbar/\piz$ ratios.   The error bars are the 
quadratic sum of statistical errors and  point-to-point systematic
errors.  There is an additional normalization uncertainty of $\sim 8\%$ 
(for $p/\pi^{+}$, $\pbar/\pi^{-}$) and  12\% (for $p/\piz$, $\pbar/\piz$),
which may shift the curves up or down, but does not affect their
shapes. In the region of overlap, the 
$\pi^{\pm}$ and $\piz$ measurements, 
with very different systematics, are consistent to within 5\% to 15\%.
 For all centralities the ratios rise steeply at low $\pt$ and
then, at a value of $\pt$ which increases from peripheral to central
collisions, level off. In central collisions the ratios are a factor of 
$\sim 3 $ larger than in peripheral events. 
At $\pt > 2$\,GeV/$c$ the peripheral Au$+$Au data agree well with 
the ratios observed in $\pp$ collisions at lower energies~\cite{ISR}
(shown with stars). The $(p+\pbar)/(\pi^{+}+ \pi^{-})$ ratio in gluon 
and quark jets produced in $e^{+}e^{-}$ collisions~\cite{DELPHI} is shown 
with dashed (dotted) line.  Above 3 GeV/$c$ the $p/\pi$, $\pbar/\pi$ 
ratios from peripheral collisions are also consistent with gluon and
quark jet fragmentation, which should be independent of the collision
system. Deviations from jet fragmentation below 3 GeV/$c$ indicate the absence
of soft hadron production in the $e^{+}e^{-}$ data.
%%%%%%%%%%%%%%%%%%%%%%%%%%%%%%%%%%%%%%%% Figure 1.
%%%%%%%%%%%%%%%%%%%%%%%%%%%%%%%%%%%%%%%%%

Hydrodynamic models have had success reproducing 
$(\pbar)p$~\cite{ppg006,star_pbar} 
and $\pi$ data~\cite{ppg006} from $\snn = 130$\,GeV Au$+$Au 
collisions~\cite{TeaneyKolb,Florkowski} and preliminary 
200\,GeV data~\cite{Rapp}. The calculations show good agreement with the 
central $p,\pbar$ and $\pi^\pm$ spectra up to $\pt \simeq $ 3 and 2 GeV/$c$,
respectively.  In peripheral collisions the calculations deviate from
the data above $\pt \simeq 1$ GeV/$c$. Within these models the large 
$\pbar/\pi$ ratio is a natural consequence of the
strong radial flow~\cite{Heinz}.  All particle spectra converge to the same
slope if $\pt$ is sufficiently larger than the particle mass $\pt \gg m_0$.  
The $\pbar/\pi$ ratio is
$R_{\pbar/\pi}\simeq 2 \exp( -\mu_b/T_{ch})$, governed only by the
baryon chemical potential $\mu_b$ and the chemical freeze-out
temperature $T_{ch}$.  Using $T_{ch}= 177$\,MeV and 
$\mu_b=29$\,MeV~\cite{PBM} $R_{\pbar/\pi}$ 
reaches a limiting value of 1.7.
Within 10\%, the same limiting behavior is
expected for all centralities, since the thermal parameters vary only 
weakly with centrality~\cite{Cleymans}. The data are not 
only below the asymptotic value but also 
show a more pronounced centrality dependence than can be accommodated 
by hydrodynamics models. This suggests that other mechanisms begin to play a 
role before the asymptotic value is reached.
At intermediate $\pt$ ($2 < \pt < 4$\,GeV/$c$), hard scattering is one 
possible mechanism that competes with ``soft'' processes 
as described by hydrodynamics. 

%%%% Comment out (TC) %%%
% Scaling behavior with respect to $\pp$ 
% collisions may help distinguish between the two. Soft processes are expected 
% to scale with the number of participant nucleons,
% $\npart$, and hard processes -  with  $\ncoll$.
% Deviations from $\ncoll$ scaling at high $\pt$ have been used 
% to establish the nuclear suppression of $\piz$ and charged
% hadrons~\cite{ppg003,ppg013,star_suppr,ppg014}.  
%%%% Comment out (TC) %%%

%%%%%%%%%%%%%%%%%%%%%%%%%%%%%%%%%%%%%%%% Figure 2.
\begin{figure}[tb]
\center
\includegraphics[width=1.0\linewidth]{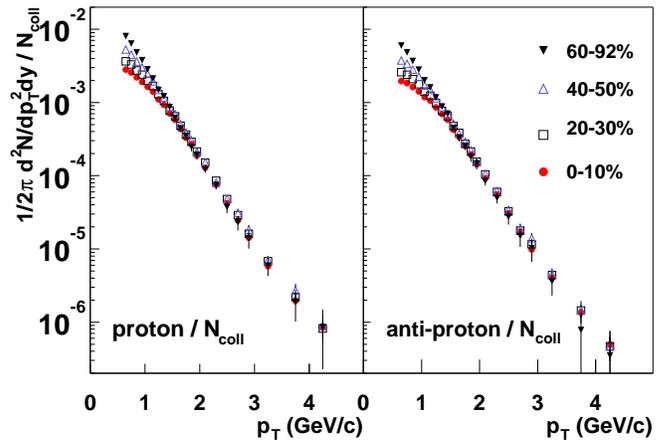}
\caption[]{$p$ and $\pbar$ invariant yields scaled by $\ncoll$.
 Error bars are statistical.
 Systematic errors on $\ncoll$ range from $\sim 10$\% for central to
 $\sim 28$\% for 60-92\% centrality. Multiplicity dependent normalization 
errors are $\sim 3\%$.}
\label{fig:2}
\end{figure}

Figure~\ref{fig:2} shows the
$p$ and $\pbar$ spectra for different centralities 
(0--10\%, 20--30\%, 40--50\%, 60--92\%) scaled by
the corresponding value of $\ncoll$~\cite{ppg014}. Error bars are statistical 
only. Multiplicity dependent systematic errors are of the order 3\%. Errors 
on $\ncoll$ range from $\sim 10\%$ for central to $\sim 28\%$ for the 
peripheral event class.  
Below $\pt \simeq 1.5$\,GeV/$c$ 
the $p$ and $\pbar$ yields scale slower than $\ncoll$ as expected 
for soft processes, and the effect of the radial flow on the shape of the
spectra is clearly visible.  The inverse slopes
gradually increase from the most peripheral to the most central event
class.  Beyond $\pt \simeq$ 1.5 GeV/$c$ all spectra converge to the
{\emph {same}} slope and seem to obey $\ncoll$ scaling as expected 
for production due to hard processes in the absence of nuclear 
effects.

%%%%%%%%%%%%%%%%%%%%%%%%%%%%%%%%%%%%%%%% Figure 3.
%\vspace{-1cm}
\begin{figure}[tb]
\center
\includegraphics[width=1.0\linewidth]{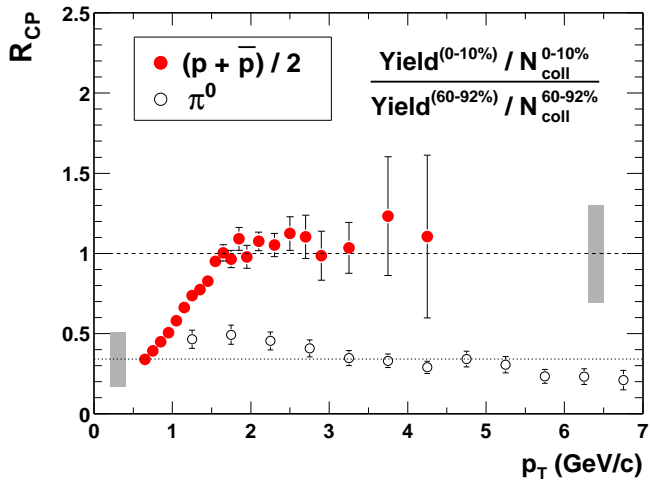}
\caption[]{Nuclear modification factor $R_{CP}$ for $(p+\pbar)/2$ 
(filled circles) and $\piz$. Dashed and dotted lines indicate $\ncoll$ and 
$\npart$ scaling; the shaded bars show the systematic errors on 
these quantities.}
\label{fig:3}
\end{figure}
%%%%%%%%%%%%%%%%%%%%%%%%%%%%%%%%%%%%%%%%%

Figure~\ref{fig:3} compares the $\ncoll$ scaled central to
peripheral yield ratios, 
\begin{equation}
R_{\rm CP} =\frac{yield^{0-10\%} / N_{coll}^{0-10\%}}{yield^{60-92\%} /  
N_{coll}^{60-92\%}}, 
\end{equation}
for $(p+\pbar)/2$ and $\piz$.
In the $\pt$ range from 1.5 to 4.5 GeV/$c$, $p$ and $\pbar$ are not
suppressed in contrast to  $\piz$ which are reduced by a factor of 2-3.
Moreover, this behavior holds for all centrality selections
(Fig.~\ref{fig:2}), while the suppression in the $\piz$ yields
increases from peripheral to central collisions~\cite{ppg014}. 
The apparent scaling with $\npart$ for $\pt \simeq 4$\,GeV/$c$, of inclusive charged  
hadrons~\cite{PHOBOS} which has been interpreted in terms of saturation 
scenario~\cite{saturation} appears to be somewhat coincidental, since we 
observe a strong species dependence not expected in the model.
However, the interpretation in terms of soft
and hard processes is also not straightforward. 
If both $\pi$ and $p$, $\pbar$ originate from the fragmentation of 
hard-scattered partons that lose energy in the medium, the nuclear modification
factor $R_{\rm CP}$ should be independent of particle species contrary
to our result.  As discussed above, for a ``hard''
description to hold, the particle ratios $\pbar/\pi$
and $p/\pi$ should reflect the fragmentation function, which favors
pion production.

It is possible that nuclear effects such as the 
``Cronin effect''~\cite{cronin1977,antreasyan1979} contribute to the observed 
large $(\pbar)p/\pi$ ratios.  In $p+$A collisions at 
energies up to $\sqrt{s} = 38.8$ GeV a nuclear enhancement 
beyond $\ncoll$ scaling has been observed for $\pi, K, p$ and their
anti-particles~\cite{Straub}. The effect is larger for $p(\pbar)$ 
than for $\pi$ which leads to an enhancement of the 
$(\pbar)p/\pi$ ratio compared to $\pp$ collisions.  For $p+W$ the 
increase is a factor of $\sim 2$ in the range 3~$<~\pt~<~6~$GeV/$c$.
Theoretical descriptions assume that the effect is due to initial state 
scattering or $\pt$-broadening~\cite{pt_broadening}.
%%%%% Replaced by PPG028 results (TC) %%%
% For RHIC energies the predicted Cronin enhancement is 
% small~\cite{Cronin_RHIC} and therefore not likely to explain the 
% large $p/\pi$,$\pbar/\pi$ ratios.
%%%%% Replaced by PPG028 results (TC) %%%
Recent results comparing charged hadrons to $\piz$ in d$+$Au
at $\snn=200$\,GeV suggest that the Cronin effect in baryons is different 
from that in mesons~\cite{ppg028}.
Another possibility is that the variation in 
the $p/\pi$ ratio with centrality reflects a medium-induced difference 
in the formation time of baryons and mesons - an effect which has been 
cited to explain DIS results~\cite{HERMES}.

Recently, the abundance of $p$ relative to $\pi$ in central
collisions has been attributed to the recombination, rather than
fragmentation, of quarks~\cite{recombination}.  
%%%% Comment out (TC) %%%%
% In this picture the
% partons from a thermalized system recombine and with the addition of
% quark momenta, the soft production of baryons extends to much larger
% values of $\pt$ than that for mesons. 
%%%% Comment out (TC) %%%%
In this model, recombination for $p$ and $\pbar$ is effective up to 
$\pt \simeq 5$\,GeV above which fragmentation dominates for all particle species.   
Another explanation of the observed large baryon content invokes a
topological gluon configuration: the baryon junction~\cite{bjunctions}.  
%%%% Comment out (TC) %%%%
% With the hard part of the spectrum largely suppressed, this production 
% mechanism is revealed and becomes dominant in the region of $\pt$ where 
% the $p/\pi, \pbar/\pi$ enhancement is observed~\cite{ptopi_bjunctions}.  
%%%% Comment out (TC) %%%%
A centrality dependence, which is in qualitative agreement with our 
results, has been predicted~\cite{ptopi_bjunctions}.
In both theoretical models, the baryon/meson enhancement is
limited to $\pt < $ 5--6\,GeV/$c$.  The identification of charged
particles beyond $\pt \approx 4.5$~GeV/$c$ is not yet
possible with the current PHENIX configuration, however 
the baryon content at high $\pt$ can be tracked
indirectly using the $h/\piz$ ratio.
%%%%%%%%%%%%%%%%%%%%%%%%%%%%%%%%%%%%%%%% Figure 4.
%\vspace{-1cm}
\begin{figure}[tb]
\center
\includegraphics[width=1.0\linewidth]{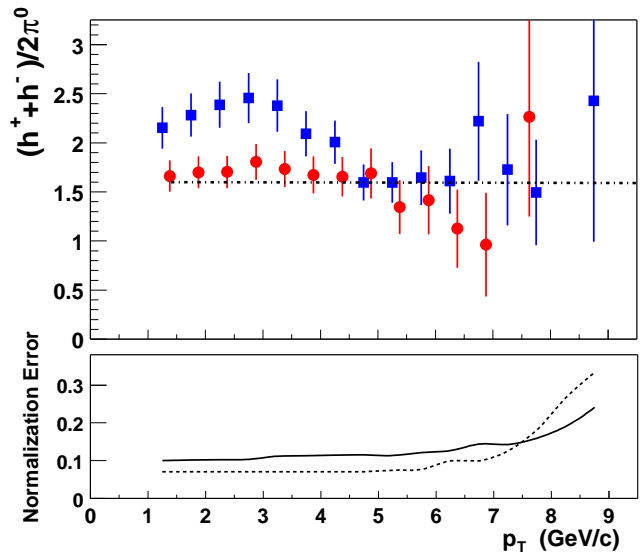}
\caption[]{Charged hadron to $\piz$ ratio in central (0-10\% -  
squares) and peripheral (60-92\% - circles) Au+Au collisions.  The peripheral 
data points are offset by +130~MeV/c for clarity.  The line at 1.6 is the 
$h/\pi$ ratio measured in $\pp$ collisions~\cite{ISR}. The lower panel shows 
fractional normalization error common to both centrality selections (solid) 
and the relative error between the two (dashed).}
\label{fig:4}
\end{figure}
%%%%%%%%%%%%%%%%%%%%%%%%%%%%%%%%%%%%%%%%%

Figure~\ref{fig:4} shows $h/\piz$ for central and peripheral Au$+$Au
collisions.  The error bars represent the quadratic sum of statistical
and point-to-point systematic errors.  
%%%% Comment out (TC) %%%%
% The solid line is the 
% fractional error common to both centralities; the dashed line shows the 
% error that can move each of the curves separately with respect to the other. 
% The line at $R_{h/\piz} = 1.6$ indicates the value measured in $\pp$ 
% collisions at lower energies~\cite{ISR}.  
%%%% Comment out (TC) %%%%
In peripheral Au+Au collisions, 
$R_{h/\piz}$ is consistent with the measurement in $\pp$.  
In central collisions in the region $1 < \pt
<$ 4.5 GeV/$c$, $R_{h/\piz}$ is enhanced by as much as 50\% above the
$\pp$ value.  As shown in Fig.~\ref{fig:1}, this enhancement is due to a
large baryon contribution. Above $\pt \simeq 5$\,GeV/$c$, the
particle composition is consistent with that measured 
in $\pp$ collisions. This indicates that the centrality-scaling of the 
$p$  yields should become consistent with that of $\pi$ at 
higher $p_T$ ($ \gtrsim 5$~GeV/$c$). Similar trends are observed in
$\Lambda$ and $K^{0}_{S}$ measurements by the STAR collaboration~\cite{star_k0}. 

We have presented a systematic study of high-$\pt$ particle composition
in Au$+$Au collisions at $\snn = 200 $\,GeV as a function of centrality.
A large $p$ and $\pbar$ contribution which increases from
peripheral to central collisions is observed in the range $ 1.5 < \pt
< 4.5 $\,GeV/$c$. In this $\pt$ range, the $p$ and $\pbar$ yields scale with 
$\ncoll$, as expected for hard-scattering.  This is in contrast 
to the centrality-dependent suppression of $\piz$ production.  The baryon 
enhancement with respect to $\pi$ seems to be limited to transverse 
momenta $\pt \sim 5$\,GeV/$c$, as
deduced from the measurement of the ratio of inclusive charged
hadrons to $\piz$.  We conclude that $\pi$ and $(\pbar)p$ 
have different dominant production mechanisms for $\pt < 5$\,GeV/$c$.

%\section{Acknowledgements}   % Run-2 short form for PRL

We thank the staff of the Collider-Accelerator and Physics
Departments at BNL for their vital contributions.  We acknowledge
support from the Department of Energy and NSF (U.S.A.), MEXT and
JSPS (Japan), CNPq and FAPESP (Brazil), NSFC (China), CNRS-IN2P3
and CEA (France), BMBF, DAAD, and AvH (Germany), OTKA (Hungary), 
DAE and DST (India), ISF (Israel), KRF and CHEP (Korea),
RMIST, RAS, and RMAE, (Russia), VR and KAW (Sweden), U.S. CRDF 
for the FSU, US-Hungarian NSF-OTKA-MTA, and US-Israel BSF.

%REFERENCES:

%FIGURES:  Place all the figures here (after the references) in sequence.

%%%\end{multicols}    % This is only used with the "multicols" style

\end{document}